# Proton elastic scattering from $^8$Li and $^9$Li nuclei within Glauber theory

© 2011 E. T. Ibraeva[1], M. A. Zhusupov[2], O. Imambekov[2]

[1]Institute of Nuclear Physics NNC RK, 050032, Ibragimov str. 1, Almaty, Kazakhstan
[2]Al-Farabi Kazakh National University, Al-Farabi av. 71, Almaty, Kazakhstan

Calculation of elastic $p^8$Li- and $p^9$Li-scattering differential cross sections, performed at two energies 0.07 and 0.7 GeV/nucleon within Glauber multiple diffraction scattering, are presented and discussed. Three-body wave functions: $\alpha-t-n$ (for $^8$Li) and $^7$Li$-n-n$ (for $^9$Li) with realistic potentials of intercluster interactions were used there. Sensitivity of elastic scattering to proton-nucleus interaction and nuclear structure has been studied. In particular, dependence of differential cross section on contribution of higher-order collisions, scattering at core and at periphery nucleons, on contribution of minor wave function components has been calculated. Comparison was made with available experimental data and with optical model calculations.

## 1. INTRODUCTION

Capability for studying nuclear matter has been expanded greatly with receipt of beams of radioactive nuclei. Measurement of differential and total cross sections for proton scattering at these nuclei within inverse kinematics provides important information on their structure: irregularity of neutron and proton (halo) density, new deformation regions and new type of collective excitation at low energy (soft dipole resonance), non-regularity in shells population.

Now we can take as a fact [1] that neutron rich isotopes $^8$Li, $^9$Li consist of core and one (for $^8$Li) and two (for $^9$Li) valence neutrons. Microscopic multicluster model predicts the skin thickness of about 0.4 fm for $^8$Li, $^9$Li [2] which leads us to the conclusion that (also supported by the data on root-mean-square (rms) radii) the $^8$Li, $^9$Li nuclei do not have intrinsic halo structure, being more like skin-nuclei, i.e. such nuclei where presence of rich neutrons do not result in considerable increase of radius but in higher concentration of neutrons in the near-surface region of the nucleus. So, the rms radius of $^6$Li (which is not a halo-nucleus) is equal to 2.44 fm and is not different from nuclei rms radii presented in the Tables I – II. Regarding spatial relation between halo neutrons, bineutron configuration of rich neutrons in $^8$Li, $^9$Li prevails over the cigar-like one, as shown in [3].

Experimental data on measuring differential cross sections (DCSs) of proton scattering from $^{8,9}$Li nuclei within inverse kinematics, had carried out at energies 0.7 GeV/nucleon [4] with IKAR facility in GSI (Darmstadt, Germany) and at energy of 0.06 GeV/nucleon (from $^9$Li) [5] in RIKEN accelerator laboratory (Japan), confirm the presence of such structure. In the both experiments the DCS of elastic scattering was measured: at momentum transfer $t = 0.0016 - 0.049$ (GeV/c)$^2$ corresponding to front scattering angles $\theta = 2 - 11°$ [4], and at the angles $\theta = 20 - 60°$ [5].

These experimental data were analyzed within various theoretical concepts. At the energies 0.06 GeV/nucleon – mainly within optical [5-10] (with the fit of the potential parameters) or folding [1, 11, 12] (with the various NN-potentials and densities) models. There the DCS could not be described correctly in all the measured angular range. To describe correctly the experimental data it is required to modify significantly the optical potential by adjusting the imaginary part [7, 8], or to multiply the real part of the folding potential by a certain normalization [12]. The best agreement with experiment was achieved when using few-particle nuclear wave functions (WFs) in the calculation, such as ones in [1], since they describe the asymptotic behavior of WFs better than most of the shell models. At the energy of 0.70 GeV/nucleon Glauber theory of multiple scattering [13] is used to calculate the DCS. The sensitivity of DCSs for different parameterizations of the nucleon density distribution [4, 14] is checked, from which conclusions about the size of the core, halo and the matter radius of the studied nuclei are drawn out.

$^8$Li nucleus is of interest not only as one-neutron halo nucleus, but also for nuclear astrophysics, since the reaction of radiative capture of $^7$Li(n,γ)$^8$Li closes the gap (for A = 8) in the chain of thermonuclear fusion reactions for CNO-elements in non-standard model of nucleosynthesis, after the so-called inhomogeneous Big Bang. Depending on the speed of this reaction, our understanding of the evolution of the universe can change. $^9$Li nucleus is of interest as an exotic neutron-rich nucleus and as the core of extensively studied $^{11}$Li nucleus, when it is considered within $^9$Li+n+n-model.

Initially, the density of $^{8,9}$Li nuclei was assessed as a single-particle density in the Gaussian or in the oscillator form, which is obviously not enough to describe all of their static characteristics, though that reproduces their empirical mean square radii.

Significant progress in numerical methods for calculating the few-particle systems has been achieved with the use of high-speed computers and the development of new counting algorithms. Modern $^8$Li nucleus WF are calculated within few-particle models in different ways: MRG with Volkov NN-potential [15] in a basis of hyperspherical harmonics with coupled interaction potentials, where the Pauli principle was accounted for by introducing a repulsive core [16], within the SMEC (Shell Model Embedded in the Continuum), where the coupling takes place between the continuous spectrum channels and bound states [17]. These WFs were mostly used for calculating the radiative capture reaction $^7$Li(n,γ)$^8$Li, studying the energy spectrum and electromagnetic charcteristics. In the three-body approach, the main effects of strong deformation, dynamic nuclear polarization and core excitation are recorded simultaneously at the same time, unlike two-body one. At the same time, four-and five-particle configurations for the $^8$Li nucleus are only the corrections to the three-particle body channel [16].

The $^9$Li WF was calculated in a 4-body α+t+n+n-model within a stochastic variational [2, 18] and quantum Monte Carlo method [19] using a realistic two-and three-body potentials: Argon (AV8, AV18), Urban IX (UIX), Illinois (Il1-Il4). These WFs reproduce all the static characteristics, the energy of the ground and nine excited states and $^9$Li+$^{12}$C reaction cross section $\sigma_{reac}$ at 800 MeV/nucleon [2]. In [4], the nuclear density of the $^{6,8,9}$Li isotopes is described by four phenomenological distributions in several models (symmetrized Fermi, Gaussian, oscillator), and all of them roughly equally reflect the matter density of $^9$Li and DCS at E = 700 MeV/nucleon. Difference between the proton and neutron radii calculated here was 0.48 fm for $^9$Li, which agrees with that predicted in the stochastic variational multicluster calculation - 0.42 fm [18] and quantum Monte Carlo method − 0.53 fm [19].

Series of DCSs calculation for elastic proton scattering at $^{8,9}$Li isotopes in Glauber multiple diffraction scattering has been presented in the present report. In contrast to the traditional calculation of characteristics within Glauber theory, where the nuclear density instead of WFs, we use the WFs obtained within the modern three-body nuclear models: α-t-n (for $^8$Li) [20-22] and $^7$Li-n-n (for $^9$Li) [23, 24]. Potentials of intercluster interactions reproduce corresponding experimental phases of elastic scattering and spectroscopic characteristics of the stated nuclei. This makes it possible to determine how structural features of nuclei appear in the characteristics of elastic scattering. Use of WFs presented in analytic form and expansion of Glauber operator into multiple scattering series in the form that can be well adjusted to the weakly-bound clusters in halo-nuclei allowed calculating scattering matrix elements analytically which increased the accuracy of calculations. We also compare different approaches to DCS assessment in order to verify the models and contribution of higher-order terms in the multiple scattering series.

In the present work the main attention is paid to the question, what particular properties of multiparticle models of halo nuclei are probed in proton elastic scattering at the intermediate (from tens up to hundreds of MeV/nucleon) energies. Our particular interest is in sensitivity of this process to the dynamics of proton-nucleus interaction, dependence of DCS on contribution of higher-order scattering and on WF minor components at various energies of incident particles.

In this paper we use the results of the earlier analysis of the WF [25, 26] and perform calculations with those of them that most adequately describe the static characteristics and with some previous versions of the WFs, to show how the improvement of the WFs affects the description of the DCSs. Comparison with available experimental data [4, 5] and calculation of other authors made within the optical model [9, 11] has been made.

## 2. BRIEF FORMALISM

According to the Glauber multiple scattering theory, amplitude of proton elastic scattering at a compound nucleus of mass A can be described (according to [13]) as an integral over the target parameter $\rho_\perp$:

$$M_{if}(\mathbf{q}_\perp) = \sum_{M_J M'_J} \frac{ik}{2\pi} \int d\boldsymbol{\rho}_\perp d\mathbf{R}_A \exp(i\mathbf{q}_\perp \boldsymbol{\rho}_\perp) \delta(\mathbf{R}_A) \langle \Psi_i^{JM_J} | \Omega | \Psi_f^{J'M'_J} \rangle \qquad (1)$$

The sign «⊥» denotes two-dimensional vectors in the plane perpendicular to the incident beam, $\langle \Psi_i^{JM_J} | \Omega | \Psi_f^{J'M'_J} \rangle$ – transition amplitude from initial $\Psi_i^{JM_J}$ to final $\Psi_f^{J'M'_J}$ state of the nucleus under the action of Ω operator; in case of elastic scattering $\Psi_i^{JM_J} = \Psi_f^{J'M'_J}$, $\mathbf{R}_A = \frac{1}{A}\sum_{n=1}^{A}\mathbf{r}_n$ – center-of-mass coordinate for the nucleus, **k** – incident particle's momentum in the c.m.s., $\mathbf{q}_\perp$ − momentum transferred in the reaction.

In the dynamic multi-cluster models [27, 28], description of a nucleus as a system of interacting clusters employs a test function in the form of the product of internal WFs of the subsystems of various cluster configurations of the particles put into relation with Jacoby coordinates **r, R**:

$$\Psi_{i,f}^{JM_J} = \Psi_1 \Psi_2 \Psi_3 \Psi^{JM_J}(\mathbf{r}, \mathbf{R}), \qquad (2)$$

where $\Psi_1$, $\Psi_2$, $\Psi_3$ − internal WFs of the clusters taken to be the same as of free particles; $\Psi^{JM_J}(\mathbf{r}, \mathbf{R})$ − WF of their relative motion. The index 1 denotes α-particle (in α−t−n- model), $^7$Li (in $^7$Li−n−n-model), index 2 − t (in α−t−n- model), n (in $^7$Li−n−n-model), index 3 − n (in α−t−n- and in $^7$Li−n−n models). Coordinate **r** describes relative α−t- (in α−t−n-model) and relative n−n- (in $^7$Li−n−n-model) motion; orbital momentum λ with the projection μ is related to it; coordinate **R** describes relative motion between centers of mass α−t- (in α−t−n-model), n−n- (in $^7$Li−n−n-model) and the remaining cluster (n, $^7$Li); orbital momentum l with the projection m is related to it. The WF for relative motion can be expanded into series in partial waves:

$$\Psi^{JM_J}(\mathbf{r},\mathbf{R}) = \sum_{\lambda lLS} \Psi^{JM_J}_{\lambda lLS}(\mathbf{r},\mathbf{R}). \tag{3}$$

Each partial function can be factorized into radial and spin-angular ones:

$$\Psi^{JM_J}_{\lambda lLS}(\mathbf{r},\mathbf{R}) = \Phi_{\lambda l}(r,R) F^{JM_J}_{\lambda lLS}(\mathbf{r},\mathbf{R}). \tag{4}$$

Radial part of WF is approximated by linear combination of Gauss functions:

$$\Phi_{\lambda l}(r,R) = r^{\lambda} R^{l} \sum_{ij} C^{\lambda l}_{ij} \exp(-\alpha_i r^2 - \beta_j R^2). \tag{5}$$

Weights of the components $C^{\lambda l}_{ij}$ can be found by numerical calculation of the Schrödinger equation using variational method; coefficients $\alpha_i$, $\beta_j$ are given in the tangential grid. The spin-angular part

$$F^{JM_J}_{\lambda lLS}(\mathbf{r},\mathbf{R}) = \sum_{M_L M_S \mu m} \langle \lambda\mu lm | LM_L \rangle \langle s_1 m_1 s_2 m_2 | SM_S \rangle \langle LM_L SM_S | JM_J \rangle Y_{\lambda\mu}(\mathbf{r}) Y_{lm}(\mathbf{R}) \chi_{SM_S} \tag{6}$$

is product of Clebsh-Gordan coefficients and spherical $Y_{\lambda\mu}(\mathbf{r})$, $Y_{lm}(\mathbf{R})$ and spine $\chi_{SM_S}$ functions. Clebsh-Gordan coefficients define the order of moment composition ($s_i m_i$ − spins and projections of the "valence" particles ($n, p$); $L$, $M_L$, $S$, $M_S$, $J$, $M_J$ − orbital, spin and total momenta of nuclei, respectively).

Configuration of WF is determined by quantum numbers $\lambda\, l\, L\, S$, available in the Tables I – II for the considered nuclei. Here we also present the potentials for pair interactions, weights of various configurations and main static characteristics of the nuclei calculated with different WFs.

According to the presentation of WF as (2), the operator $\Omega$ can be written in the form of expansion into series for proton multiple scattering at each incident subsystem:

$$\Omega = \Omega_1 + \Omega_2 + \Omega_3 - \Omega_1\Omega_2 - \Omega_1\Omega_3 - \Omega_2\Omega_3 + \Omega_1\Omega_2\Omega_3. \tag{7}$$

Here, subscripts 1, 2, 3 correspond to the same clusters as in WF. Calculation of a matrix element with three-body WFs (2) has been considered in details in [24−26]. Let us note that WF expanded into Gaussoids (5) and the operator $\Omega$ taken in the form (7) as in three-body WF make it possible to integrate the amplitude (1) analytically therefore improving precision of the calculations.

Differential cross section is determined as a square modulus of the matrix element:

$$\frac{d\sigma}{d\Omega} = \frac{1}{2J+1} |M_{if}(\mathbf{q}_\perp)|^2. \tag{8}$$

Let us write down several equations which we will need while discussing the results. Putting $^7$Li$-n-n$-model WF of $^9$Li nucleus into (1): $\Phi^{JM_J}_{\lambda lLS} = \Psi_{7Li}(\mathbf{R}_{7Li}) \varphi_n(\mathbf{r}_1) \varphi_n(\mathbf{r}_2) \Psi^{JM_J}_{\lambda lLS}(\mathbf{r},\mathbf{R})$ - and squaring the matrix element in Eq. (8), we get the p$^9$Li-scattering DCS which accounts for three WF components (see the Table II, versions 2 and 3):

$$\frac{d\sigma}{d\Omega} = \frac{1}{2J+1} \cdot \left| \frac{ik}{2\pi} \sum_{M_s M_s'} \int d\boldsymbol{\rho}_\perp d\mathbf{R}_9 \exp(i\mathbf{q}_\perp \boldsymbol{\rho}_\perp) \delta(\mathbf{R}_9) \left\{ \langle \Phi^{JM_J}_{000\,3/2} | \Omega | \Phi^{JM_J'}_{000\,3/2} \rangle + \right. \right.$$
$$\left. \left. + \langle \Phi^{JM_J}_{111\,3/2} | \Omega | \Phi^{JM_J'}_{111\,3/2} \rangle + \langle \Phi^{JM_J}_{111\,1/2} | \Omega | \Phi^{JM_J'}_{111\,1/2} \rangle \right\} \right|^2, \tag{9}$$

where the first, the second and the third terms within braces determine the contribution of WF components with quantum numbers ($\lambda$, l, L, S = 0, 0, 0, 3/2; 1, 1, 1, 3/2; 1, 1, 1, 1/2) accordingly.

Putting operator $\Omega$ into matrix element (1) as the series (7) and squaring it (8), we get DCS's dependence on different collision multiplicity:

$$\frac{d\sigma}{d\Omega} = \frac{1}{2J+1} \left| M^{(1)}_{ij}(\mathbf{q}) - M^{(2)}_{ij}(\mathbf{q}) + M^{(3)}_{ij}(\mathbf{q}) \right|^2, \tag{10}$$

where

$$M^{(1)}_{if}(\mathbf{q}_\perp) = \frac{ik}{2\pi} \sum_{M_s M_s'} \int d\boldsymbol{\rho}_\perp d\mathbf{R}_A \exp(i\mathbf{q}_\perp \boldsymbol{\rho}_\perp) \delta(\mathbf{R}_A) \left\{ \langle \Psi^{JM_J}_{\lambda lL} | \Omega_1 + \Omega_2 + \Omega_3 | \Psi^{JM_J'}_{\lambda lL} \rangle \right\}, \tag{11}$$

$$M^{(2)}_{if}(\mathbf{q}_\perp) = \frac{ik}{2\pi} \sum_{M_s M_s'} \int d\boldsymbol{\rho}_\perp d\mathbf{R}_A \exp(i\mathbf{q}_\perp \boldsymbol{\rho}_\perp) \delta(\mathbf{R}_A) \left\{ \langle \Psi^{JM_J}_{\lambda lL} | \Omega_1\Omega_2 + \Omega_1\Omega_3 + \Omega_2\Omega_3 | \Psi^{JM_J'}_{\lambda lL} \rangle \right\}, \tag{12}$$

$$M_{if}^{(3)}(\mathbf{q}_\perp) = \frac{ik}{2\pi} \sum_{M_s M_s'} \int d\boldsymbol{\rho}_\perp d\mathbf{R}_A \exp(i\mathbf{q}_\perp \boldsymbol{\rho}_\perp)\delta(\mathbf{R}_A)\left\{ \left\langle \Psi_{\lambda l L}^{JM_J} \middle| \Omega_1 \Omega_2 \Omega_3 \middle| \Psi_{\lambda l L}^{JM_J'} \right\rangle \right\}. \quad (13)$$

Here, $M_{if}^1(\mathbf{q}_\perp)$, $M_{if}^2(\mathbf{q}_\perp)$, $M_{if}^3(\mathbf{q}_\perp)$ – partial amplitudes of single-, double- and triple collisions.

## 3. RESULTS AND DISCUSSION

Upon calculation of DCSs for elastic scattering of protons from $^8$Li, $^9$Li nuclei in Glauber formalism we compare them with available experimental data and calculation made within the optical model [9, 11].

Fig. 1 presents calculated DCS at $E = 0.060$ (*a*) and $E = 0.698$ GeV/nucleon (*b*) for $p^8$Li-scattering with WF in the $\alpha$–$t$–$n$-model and different potentials of intercluster $\alpha$−$t$-interaction described in the Table I. Curves *1, 2, 3* are calculated with WFs variants 1, 2 and 3. For both energies all three curves, equally describing DCSs at small scattering angles, are quite different only in the region of their minima. At small scattering angles the momentum transferred is low and only a periphery of nucleus can be probed (i.e. asymptotic behavior of WF). Behavior of three-particle WF at asymptotic is almost the same, so the DCSs for all variants of WF calculations are the same at small angles. At larger scattering angles the transferred momentum increases, particles interact mostly in the inner part of the nucleus where particle correlation effects (which, actually, make the models different from each other) are stronger and one can observe differences in angular distributions. The region with minimal DCSs is the most sensitive one. It is known [35] that scattering at nuclei with large quadrupole momenta demonstrates DCSs with much less expressed diffraction picture than that from spherically symmetrical ones. Let us compare this factor for the curves *1 – 3*. Curve *1*: $Q = 16.55$ mb, contribution of the components with $L = 2$ equals to zero (the Table I, version 1), curve *2*: $Q = 18.94$ mb, total contribution of the components with $L = 2$ equals to $P = 0.010$ (the Table I, version 2), curve *3*: $Q = 30.36$ mb, contribution of the components with $L = 2$ equals to $P = 0.109$ (the Table I, version 3). There is an ambiguous situation with the experimental quadrupole momentum: the review [34] presents the value $Q=24(2)$ mb, while the new measurement techniques increased it up to $Q = 32.7(6)$ mb [29], so the quadrupole momentum calculated in [20] is almost two times less than the experimental one. Trying to improve these results, in [21] there was made an attempt to increase the number of configurations taken into consideration (particularly, the configuration with $\lambda = 3$ was included since this particular state demonstrates a low-energy resonance in $\alpha t$-system), and calculation was made with additional $\alpha t$-potentials that influences the properties of the $^8$Li basic state more than $\alpha n$- or $tn$-ones. Still, the weights of considered configurations turned out to be too low (by two orders of magnitude less than that of the prevailing component $\lambda lLS=1111$) to influence considerably the calculation, so this action only resulted in minor increase of the quadrupole momentum (version 2 in the Table I). Only the last paper [22] considered tensor forces in the interaction potential and achieved acceptable (within 10%) agreement of the calculated quadrupole momentum with the experimental one (version 3 in the Table I). Though tensor forces are small compared to the central ones, they lead to mixing over orbital momentum increasing contribution of WF minor components which, in their turn, determine such characteristics of nuclei as quadrupole and magnetic moments. The Table I also shows that the tensor forces, including contribution of configurations with $\lambda lL=112$, improve $Q$ considerably. Presented data make it obvious that filling the minimum of the cross section is proportional to the weight of quadrupole component in the nucleus WF. With WF calculated under version 3 magnetic moment is also described adequately.

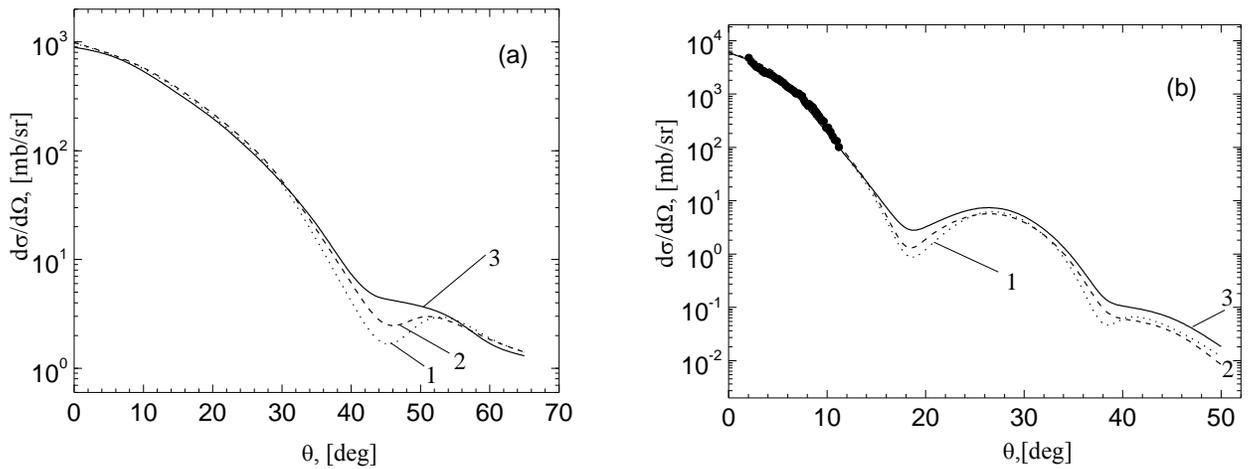

Fig. 1. Differential cross section of $p^8$Li-scattering as a function of model WFs calculated with different variants of intercluster potentials for $\alpha$–$t$-interactions at $E = 0.060$ (left panel) and $E = 0.698$ GeV/nucleon (right panel). Curves *1, 2, 3* represent calculations under WF versions 1, 2 and 3 as in the Table I. Experimental data at this and other figures at $E = 0.698$ GeV/nucleon are taken from [4].

TABLE I. Interaction potentials, included wave function configurations and static characteristics of the $^8$Li nucleus in the $\alpha-t-n$ model that were obtained by using them

| Potential | Version 1 [20] | Version 2 [21] | Version 3 [22] |
|---|---|---|---|
| $\alpha-t$ | Gaussian potential containing eight parameters | Gaussian potential in the Buck form with a supersymmetric repulsive part at short distances | The same as under version 2, but with the inclusion of a tensor interaction |
| $\alpha-n$ | Gaussian potential split in the parity of the orbital angular momentum | The same as under version 1 | The same as under version 1 |
| $t-n$ | Gaussian potential split in the parity of the orbital angular momentum | The same as under version 1 | The same as under version 1 |
| Configuration $\lambda$ $l$ $L$ $S$ | Weight of the configuration (P) | | |
| 1  1  1  1 | 1.00 | 0.9899 | 0.8854 |
| 1  1  2  1 |  | 0.0024 | 0.0768 |
| 3  1  2  1 |  | 0.0045 | 0.0378 |
| 3  1  2  0 |  | 0.0032 |  |
| Static features |  |  |  |
| $R_{rms}$, fm $R_{rms}^{\exp}$ = 2.50(6) fm [4] | 2.36 | 2.348 | 2.38 |
| $E$, MeV $E_{exp.} = -4.501$ MeV [29] | -3.82 | -4.406 | -4.657 |
| $\mu$, $\mu_0$ $\mu_{\exp} = 1.65\mu_0$ [30] | 1.48 | 1.408 | 1.624 |
| $Q$, mb $Q_{exp.}$ = 32.7(6) mb [29] | 16.55 | 18.94 | 30.36 |

    Experimental data for are only available for $E = 0.698$ GeV/nucleon [4]; comparison to these data is shown in Fig. 1a. At small scattering angles below $\theta \sim 11.3°$ (that corresponds to $t = 0.049$ (GeV/c)$^2$) the agreement with the experiment is observed for all variants of our calculation. Unfortunately, experiment does not reach the minimum region $\theta \sim 20°$ where the DCS is sensitive to WF behavior.

    Fig. 2 presents DCS for $p^9$Li-scattering with different WFs models at $E = 0.06$ (a) and $E = 0.703$ GeV/nucleon (b). The lines *1, 2, 3* represent our calculation with WF in $^7$Li$-n-n$-model under different versions (1, 2 and 3, respectively) of intercluster interactions in the Table II. Experimental data is from [5] for $E = 0.06$ and from [4] for $E = 0.703$ GeV/nucleon. Fig. 2 (a) reports on calculation made within optical model by [11] − curve *4* and by [9] − curve *5*.

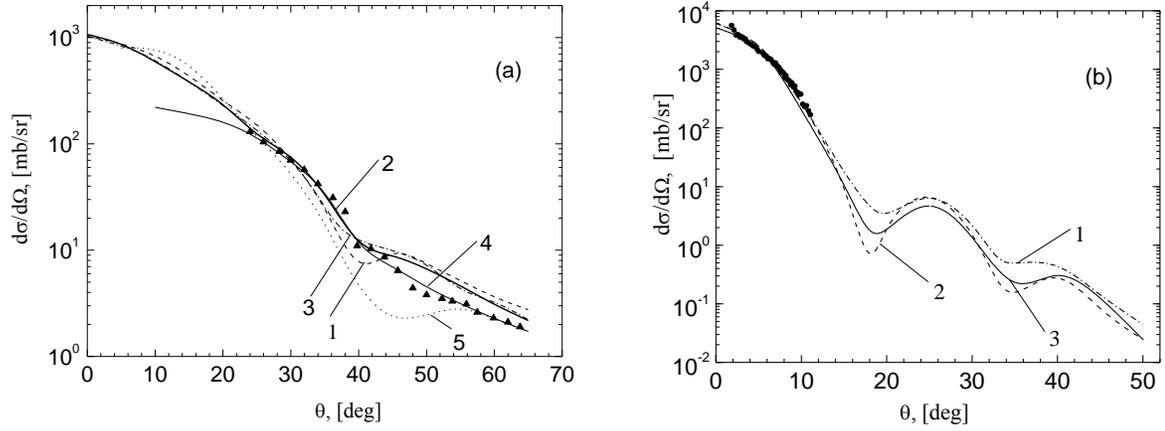

Fig. 2. Differential cross section of $p{}^9Li$-scattering as a function of model WFs calculated with different variants of intercluster interactions at $E$ = 0.06 (left panel) and 0.717 GeV/nucleon (right panel). Lines *1*, *2*, *3* represent calculations under WFs versions 1, 2 and 3 as in the Table II. Experimental data at this and other figures at $E$ = 0.06 GeV/nucleon are taken from [5], at $E$ = 0.717 GeV/nucleon – from [4]. Line *4* at the left panel is taken from [11], line *5* – from [9].

TABLE II. Interaction potentials, included wave function configurations and static characteristics of the ${}^9Li$ nucleus in the ${}^7Li-n-n$-model that were obtained by using them

| Potential | Used variant | | |
|---|---|---|---|
| ${}^7Li-n$ | Deep attractive potential involving forbidden states and having the Buck form without exchange terms | Deep attractive potential involving forbidden states and having the Buck form with exchange terms featuring a strong spin dependence | |
|  | Version 1 [24, 25] | Version 2 [24, 25] | Version 3 [24, 25] |
| $n-n$ | Hasegava-Nagata potential [31] for odd waves, Afnan-Tang potential [32] with a repulsive core for even waves[a] | The same as under version 1 | The same as under version 1[b] |
| Configuration $\lambda$ $l$ $L$ $S$ | Weight of the configuration (P) | | |
| 0  0  0    3/2 | 0.051 | 0.984 | 0.662 |
| 1  1  1    3/2 | 0.471 | 0.015 | 0.169 |
| 2  2  1    3/2 | 0.246 |  |  |
| 1  1  1    1/2 | 0.097 | 0.001 | 0.169 |
| 2  2  1    1/2 | 0.050 |  |  |
| 3  3  1    3/2 | 0.071 |  |  |
| 3  3  1    1/2 | 0.014 |  |  |
| Static features |  |  |  |
| $R_{rms}$, fm $R_{rms}^{\exp}$ = 2.44(6) fm [4] | 2.38 | 2.40 | 2.46 |
| $E$, MeV $E_{exp.}$ = − 6.096 MeV [33] | -9.01 | -6.20 | -5.906 |
| $\mu$, $\mu_0$ | 0.94 | 1.31 | 1.33 |
| $\mu_{exp} = 3.44\mu_0$ [33, 34] | | | |
| $Q$, mb $Q_{exp.} = -27.4(1)$ mb [33] | -40.0 | -23.98 | -27.93 |

[a)] Parameters potentials No. 1 (Table II) [31] and S1 (Table I) [32]
[b)] Parameters potentials No. 1 (Table II) [31] and S3 (Table I) [32]

Let's see what is the difference between various WFs versions. As it is seen from the Table II, deep attractive $^7$Li-n-potential without exchange terms, and n-n-potential of Hasegawa-Nagata and Afnan-Tang (version 1) do not reproduce the static characteristics of $^9$Li nucleus. One way to improve the three-body model is taking into account the mixing of configurations in the nucleus (as shown by the calculation for the $^8$Li nucleus, where the tensor forces were taken into account in the interaction potential). So the next step was to change the $^7$Li-n-potential: it incorporated the exchange terms with the strong spin dependence (versions 2 and 3). $\lambda\ell LS = 0\ 0\ 0\ 3/2$ is the basic configuration in this case. The calculations performed with the search for configurations, other than the basic, showed that the contribution of $\lambda\ell LS = 1\ 1\ 1\ 3/2$ and $\lambda\ell LS = 1\ 1\ 1\ 1/2$ configurations would be the next. However, a small admixture of $\lambda\ell LS \neq 0$ states with the basic one does not give the correct value of the quadrupole moment (version 2). Proper weight distribution for the main and additional configurations was found in the version 3, which led to the value of the quadrupole moment consistent with the experimental one.

Comparing our calculation with the experiment in Fig. 2 (a) we would like to note that calculation within the Glauber theory with all WFs versions agree numerically with the experimental data at the forward scattering angles only (less than $\theta < 30°$), further up to $\theta < 40°$ calculated DCS lie below the experimental ones and at $\theta > 45°$ they lie above the experimental ones. This can be explained by two factors: inapplicability of Glauber formalism to moderate and large scattering angles and poor description of nuclear interiors. The last is based on the fact that large scattering angles correspond to large transferred momenta since $q=2k\sin(\theta/2)$ that in the WF coordinate space corresponds to small relative distances between clusters and nucleons, i.e. to internal regions of a nucleus. Curve *4* from [11] has been calculated within the optical model with complex folding-potential and with the density which considered effects of nuclear medium. This curve reconstructs correctly the experimental data for the whole range of angles since the calculations are made within the optical model without any restriction for large scattering angles imposed in Glauber theory. In contrast, curve *5* of [9], where calculations were also performed within the optical model with Paris-Hamburg nonlocal potential, lies below the experimental data for almost all angles except several points in the range of angles $\theta > 55°$.

In Fig. 2 (b) calculation with WFs in the models 1, 2 and 3 reproduces DCS equally in the range of forward angles up to $\theta < 15°$. The difference takes place in the regions of diffraction minima and at large scattering angles. Let us see how DCS minimum and quadrupole moment correlate. Comparison of curves at the figure and the values of $Q$ in the Table II show that the smaller (smoother) cross section minima are, the higher quadrupole moment is. So, curve *2* corresponds to $|Q| = 23.98$ mb, curve *3* – $|Q| = 27.4$ mb, curve *1* – $|Q| = 40.0$ mb. Herewith, the most close to the experiment value was achieved with the WF under version 3. In this case, filling the minimum is related to the WF components with $L = 1$ (the Table II). The same WF brought us the values of rms radius and binding energy close to the experimental ones. Magnetic moment is not reproduced under all three versions, that indicates insufficient contribution of internal regions, while under the version 3 the WF is calculated with the Buck's attractive potential where the WF does not die out inside nucleus like in the potential with repulsive core.

All the calculations support the conclusion that DCS is just slightly sensitive to WF behavior at asymptotic (that corresponds to small scattering angles), but to larger extent depends on internal regions of nucleus, i.e. on core. Low sensitivity of elastic scattering DCS to various density distributions at small scattering angles was also discussed in [36] where DCS of elastic $p^8$He-scattering has been calculated with two different methods: JLM (Jeukenne-Lejeune-Mahaux) and eikonal approach. At small scattering angles DCSs in both approaches describe the experimental data equally well, while at large angles the calculated curves are different indicating differences in scattering at core and at skin. At the same time the authors point out that the difference in density distribution in core and at periphery is not striking and appears to be model-dependent and one needs precise cross section measurement at large angles to use it for density distribution measurement that represents a challenging task due to smallness of DCS.

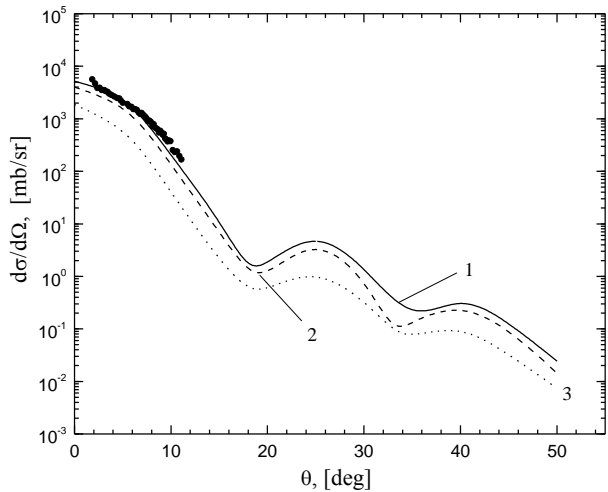

Fig. 3. Differential cross section as a function of contribution of different WF components. $p^9$Li-scattering at $E = 0.717$ GeV/nucleon. Curve *2* represents contribution of the term ($\lambda, l, L, S = 0, 0, 0, 3/2$), curve *3* – from the terms ($\lambda, l, L, S = 1, 1, 1, 3/2; 1, 1, 1, 1/2$), curve *1* – sum of all three components, same as curve *3* in Fig. 2 (right panel)).

Let us consider the contribution of various WF components to the DCS in more details using $p^9$Li-scattering as an example. Calculation made with the equation (9) with WF under the version 3 is illustrated at Fig. 3 for $E = 0.7$ GeV/nucleon: curve *2* – contribution of the first term of the eq. (9), curve *3* – of the second and third terms of the eq. (9) and, as it is seen in the figure, these curves are similar in shape and different in their absolute values only. The curve *1* is a sum of all three components (the same as the curve *3* in Fig. 2 (b)). Absolute value of the component contribution to DCS is determined by their weights (the Table II): the first WF component (curve *2*) provides main contribution to the cross section since its weight is 0.662; weight of each of two other components comprises 0.169 that is why their contribution is less (curve *3*). Still, these components provide their contribution over the whole range of angles and we need to take them into account anyway. Analysis of geometric configuration of WF components ($\lambda, l, L, S = 0, 0, 0, 3/2$), ($\lambda, l, L, S = 1, 1, 1, 3/2$) and ($\lambda, l, L, S = 1, 1, 1, 3/2$) showed that they all have two peaks of density [24]. One maximum is located closer to the centre of the nucleus, it corresponds to the configuration where three particles ($^7$Li, n, n) are located at vertices of isosceles triangle. The second maximum is located further away from the core, its configuration is similar to the "cigar-like" one (two nucleons on both sides of the $^7$Li core). The difference of geometric shapes of all three configurations is insignificant, and therefore their contributions to the DCS are similar.

Fig. 4 presents contribution of different scattering multiplicities calculated for $p^8$Li-scattering ($\Omega_1 = \Omega_\alpha, \Omega_2 = \Omega_t, \Omega_3 = \Omega_n$). The curves *1, 2* and *3* describe individual contribution of the first, second and third terms in (10). Their sum is the curve *4* – the same as the curve *3* in Fig. 1 (b). In the figure one can see that the main contribution at small scattering angles ($\theta < 20°$) is provided by single scattering with $\alpha$-, $t$-clusters and valence neutron, but its amplitude decreases rapidly and the higher multiplicities dominate at larger angles. One can also see that at small scattering angles the curve *1* lies above the summary curve *4* since in the series (10) the double scattering term is subtracted from the single one decreasing the summary DCS to be less than the DCS of single scattering. In the point where the curves of single- and double-scattering cross each other the curve *4* has its minimum due to destructive interference at squaring of matrix elements in the formula (10). After the interference minimum at $\theta \approx 18°$ double collisions begin dominating, which mainly contribute to the second cross section maximum. When $\theta \approx 39°$, the magnitudes of partial cross sections for double and triple collisions are comparable, and second minimum appears in the DCS due to their interference. When $\theta > 39°$ decisive contribution to the DCS is given by triple collision. Thus, expansion (7) provides us with a convenient way of finding a relation between the terms of single and higher-order scatterings. From the figure one can see that contribution of higher-order collisions are to be taken into consideration for adequate description of DCS in a wide range of angles (up to $\theta \sim 50°$). This idea has been supported in a range of other papers [37, 38] where contribution of higher-order terms of multiple proton scattering from $^6$He, $^{11}$Li, $^{11}$Be nuclei to cross sections have been considered.

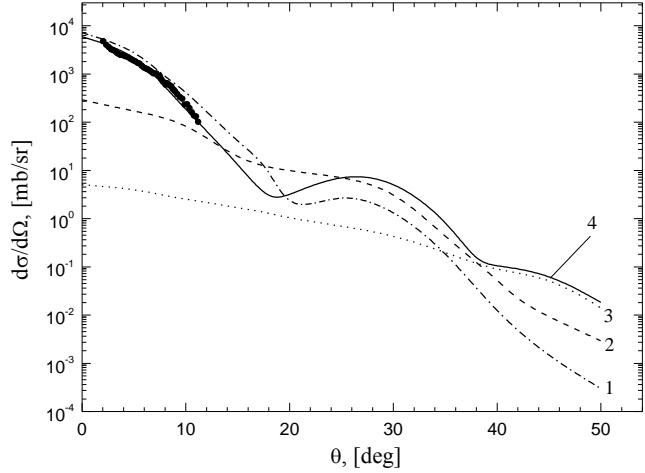

Fig. 4. Contribution of different scattering multiplicities at the operator $\Omega$ into differential cross section. Curves *1*, *2*, *3* and *4* represent contribution of single, double, triple scatterings and their summary contribution taking interference into account; $p^8$Li-scattering at E = 0.698 GeV/nucleon.

Based on the examples considered above, one can make conclusion that higher-order dynamic contribution is considerable and needs to be accounted when considering scattering at all studied nuclei in case of large transferred momentum.

There are several components in case of single scattering (11) in our model. The first one is determined by core ($\Omega_1 + \Omega_2 = \Omega_\alpha + \Omega_t$ for $^8$Li, $\Omega_1 = \Omega_{7Li}$ for $^9$Li), the second is valence term due to scattering at neutrons ($\Omega_2 + \Omega_3 = \Omega_{n1} + \Omega_{n2}$ for $^9$Li, $\Omega_3 = \Omega_n$ for $^8$Li). Therefore important information on core and skin structure is provided by single scattering.

Contribution of collisions with core and spin to single scattering DCS for $p^8$Li-scattering at $E = 0.698$ GeV/nucleon is shown in Fig. 5. Here, the curve *2* represents the summary scattering at $\alpha$ and $t$ (contribution of the first two terms in (11)), the curve *3* – scattering at nucleon (contribution of the last term in (11)), the curve *1* – the sum of all three terms in (11), the same as the curve *1* in Fig. 4. From Fig. 5 one can conclude that DCS of scattering at core (curve *2*) prevails in the entire range of angles and has structure with small minimum at $\theta = 20°$; DCS of scattering at single nucleon takes form of steadily decreasing function which magnitude is by two orders less than DCS of scattering at core even at $\theta = 30°$. Obviously, the excess neutron is localized in the surface area, and its contribution to DCS is distinguishable at small angles only. Similar results were obtained for $p^9$Li-scattering at $E = 0.703$ GeV/nucleon.

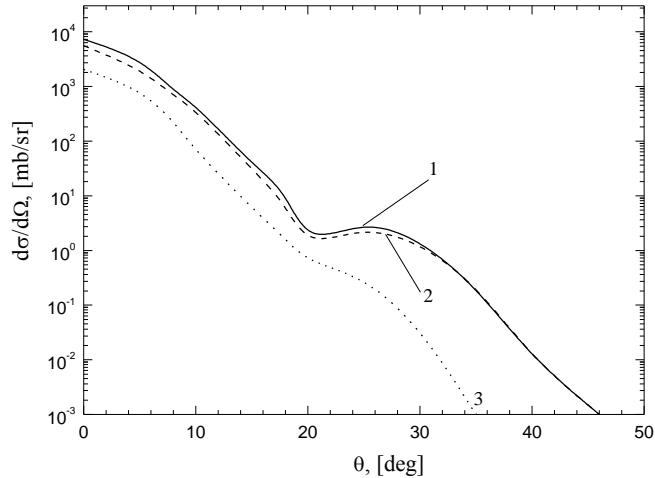

Fig. 5. Contribution of core scattering (curve *2*) and skin scattering (curve *3*) to differential cross section of $p^8$Li single scattering (curve *1*) at $E = 0.698$ GeV/nucleon.

In papers [12, 36, 39, 40] at discussing different mechanisms of reaction with halo nuclei and information that could be obtained from the measured characteristics the conclusion was made that core plays more important role than halo in description of DCS. Still partial contribution to the DCS from individual structural components of nucleus were not calculated in those works while in the present work it has been directly shown that DCS of single collisions is fully determined by scattering at core.

## CONCLUSIONS

In the present work characteristics of elastic $p^8$Li- and $p^9$Li-scattering at two energies 0.07 and 0.7 GeV/nucleon were calculated within Glauber formalism which input parameters are elementary $pN$-, $p\alpha$-amplitudes and WFs of nuclei. It is substantial in the presented calculations that we used realistic three-particle WFs obtained within the modern nuclear models. Performed calculations and analysis of DCS for elastic scattering of protons from $^8$Li, $^9$Li nuclei in inverse kinematics enable us to make the following conclusions.

1. Analyzing DCSs calculated with different model WFs both in Glauber approach and in other formalisms we have shown that DCS depends slightly on WF behavior at asymptotic; the dependence on internal part of WF is much higher. This conclusion is supported by calculating contribution of different components to the cross section of single scattering. Identifying there the terms sensitive to scattering at core and at skin, we have shown that DCS of elastic scattering at $^{8,9}$Li nuclei is mainly determined by scattering at core over the whole range of angles while scattering at periphery neutrons contributed slightly to the cross section at small angles only, since the low-density skin cannot deflect a particle to a large angle.
2. Expanding Glauber multiple scattering operator into a series of scattering at clusters and nucleons in nucleus, we calculated DCS taking into account all multiplicities and partial (single-, double- and triple) cross sections, and we showed that, while main contribution to DCS at small transferred momenta is provided by single collisions, dynamic contribution of higher order are considerable and are to be accounted at large transferred momenta.
3. Contribution of minor WF components related to tensor interactions in intercluster potentials enables us to describe adequately the quadrupole momenta of nuclei. There is a correlation between filling in the DCS minimum and contribution of minor WF components: ($\lambda$, $l$, $L$, $S$ = 1, 1, 2, 1; 3, 1, 2, 1) for $^8$Li and ($\lambda$, $l$, $L$, $S$ = 1, 1, 1, 3/2; 1, 1, 1, 1/2) for $^9$Li.
4. Comparison of calculation performed within optical model and with various model WFs showed good accuracy of Glauber approach and consistent description of the experimental data.